\documentclass
[superscriptaddress,secnumarabic,amssymb,amsmath,nobibnotes,aps,prd,showkeys,showpacs,nofootinbib,onecolumn,12pt]{revtex4}%
\usepackage{graphicx}
\usepackage{subfigure}
\usepackage{epsf}
\usepackage{bm}
\usepackage{amsmath}
\usepackage{amsfonts}
\usepackage{amssymb}%
\providecommand{\U}[1]{\protect\rule{.1in}{.1in}}
\usepackage{hyperref}
\hypersetup{
    colorlinks=true,
    linkcolor=red,
    citecolor=blue,
    filecolor=magenta,
    urlcolor=cyan,
    pdftitle={A Perturbative Analysis of Interacting Scalar Field Cosmologies},
    pdfpagemode=FullScreen}

\newcommand{\be}{\begin{equation}}
\newcommand{\ee}{\end{equation}}

\newcommand{\mincir}{\raise
-3.truept\hbox{\rlap{\hbox{$\sim$}}\raise4.truept\hbox{$<$}\ }}
\newcommand{\magcir}{\raise
-3.truept\hbox{\rlap{\hbox{$\sim$}}\raise4.truept\hbox{$>$}\ }}

\usepackage{color}

\begin{document}
\title{Interacting dark energy in curved FLRW spacetime from Weyl Integrable Spacetime}
\author{S. Chatzidakis}
\affiliation{Applied Technology High School, PO Box 66866, Al Ain, United Arab Emirates}
\author{A. Giacomini}
\email{alexgiacomini@uach.cl}
\affiliation{Instituto de Ciencias F\'{\i}sicas y Matem\'{a}ticas, Universidad Austral de
Chile, Valdivia 5090000, Chile}
\author{P.G.L. Leach}
\affiliation{Institute of Systems Science, Durban University of Technology, PO Box 1334,
Durban 4000, South Africa}
\author{G. Leon}
\email{genly.leon@ucn.cl}
\affiliation{Departamento de Matem\'{a}ticas, Universidad Cat\'{o}lica del Norte, Avda.
\ Angamos 0610, Casilla 1280 Antofagasta, Chile}
\affiliation{Institute of Systems Science, Durban University of Technology, PO Box 1334,
Durban 4000, South Africa}
\author{A. Paliathanasis}
\email{anpaliat@phys.uoa.gr}
\affiliation{Instituto de Ciencias F\'{\i}sicas y Matem\'{a}ticas, Universidad Austral de
Chile, Valdivia 5090000, Chile}
\affiliation{Institute of Systems Science, Durban University of Technology, PO Box 1334,
Durban 4000, South Africa}
\author{Supriya Pan}
\email{supriya.maths@presiuniv.ac.in}
\affiliation{Department of Mathematics, Presidency University, 86/1 College Street, Kolkata
700073, India}

\begin{abstract}
In the present article, we show that a simple modification to the Einstein-Hilbert action can explain the possibility of mutual interaction between the cosmic fluids. That is achieved considering the Weyl Integrable Spacetime in the background of a nonflat Friedmann-Lemna\^{i}tre-Robertson-Walker geometry for the universe. We show that widely-known phenomenological interacting cosmological scenarios can naturally appear in this context. We then performed the dynamical system analysis of the underlying cosmological scenario and explored many possibilities extracted from this gravitational theory.
\end{abstract}
\date{\today}
\maketitle

\section{Introduction}
\label{sec1}

The detection of the late-time accelerating phase of the universe \cite{SupernovaSearchTeam:1998fmf, SupernovaCosmologyProject:1996grv} is one of the significant discoveries in cosmology which abruptly changed our ideas on the evolution of our universe. To describe this accelerating expansion, usually two well-known approaches are considered. One is the introduction of some hypothetical dark energy (DE) fluid in Einstein's General Relativity (GR), and the other is to either modify the Einstein's GR or introduce new gravitational theories. In the latter approach, the additional (new) geometrical terms appearing from the modified (new) gravitational theory mimic the role of a DE. Following both the approaches, several cosmological models have been proposed in the last several years to explain the dynamics of the universe  \cite{Copeland:2006wr, Nojiri:2006ri, Sotiriou:2008rp, DeFelice:2010aj, Clifton:2011jh, Cai:2015emx, Nojiri:2017ncd}. Among these models, the $\Lambda$-Cold Dark Matter ($\Lambda$CDM) cosmological model, designed in the context of GR, has been quite successful in describing a large span of astronomical and cosmological data.  However, the physics of the primary two ingredients in the $\Lambda$CDM paradigm, namely, the cold dark matter and cosmological constant is not yet clearly understood despite many astronomical missions. Apart from that, the cosmological constant problem \cite{Weinberg:1988cp}, and the coincidence problem \cite{Zlatev:1998tr} require explanation in this cosmological scenario. These are not the only limitations in this successful cosmological model.
It has been  consistently observed that some key cosmological parameters estimated from the early time measurements by
Planck within this minimal $\Lambda$CDM cosmology \cite{Planck:2018vyg} significantly differ from  other astronomical missions \cite{Riess:2021jrx,Hildebrandt:2016iqg,Joudaki:2017zdt,DES:2017myr,Hildebrandt:2018yau}. The most statistically significant discrepancy  (at $5\sigma$) is the Hubble constant tension between Planck \cite{Planck:2018vyg} and Supernovae $H_0$ for the Equation of State (SH0ES) collaboration~\cite{Riess:2021jrx} which demands a revision of the $\Lambda$CDM cosmology. On the other hand, the tension in the growth of structures, quantified through the parameter $S_8$, defined as a combination of the amplitude of the matter power spectrum $\sigma_8$ with the matter density at present $\Omega_m$ ($S_8=\sigma_8\sqrt{\Omega_m/0.3}$) between Planck and cosmic shear measurements \cite{Hildebrandt:2016iqg, Joudaki:2017zdt, DES:2017myr, Hildebrandt:2018yau} has been another issue which has further strengthened the revision of the $\Lambda$CDM cosmology. The cosmological discrepancies have been one of the hot topics at present moment which fueled the scientific community to find an alternative to the $\Lambda$CDM cosmology, see Refs.
\cite{DiValentino:2020zio,DiValentino:2020vvd,DiValentino:2021izs,Perivolaropoulos:2021jda,Abdalla:2022yfr} which include a variety of cosmological models attempting to explain these discrepancies.
However, despite many attempts performed by a large number of investigators (see again  \cite{DiValentino:2020zio,DiValentino:2020vvd,DiValentino:2021izs,Perivolaropoulos:2021jda,Abdalla:2022yfr}), the tension free cosmological model is yet to be discovered. In order to be more transparent in this direction, it should be mentioned that the simultaneous solutions to both the tensions is hard to achieve. Due to the existing correlation between $H_0$ and $S_8$, the solution of the $H_0$ tension in a specific cosmological model may worsen the tension in the $S_8$ parameter in the same cosmological model. Moreover, the solution to the Hubble $H_0$ tension in a cosmological model can increase the tension in the sound horizon $r_d$ derived from the cosmic microwave background, and other cosmological probes \cite{Arendse:2019hev}. As the full cosmic picture is yet to be discovered, there is certainly no reason to favor any cosmological model over the others.  Motivated by this,  in the present work we investigate a cosmological dynamics which arises from a generalized Einstein-Hilbert action, and interestingly, this offers a novel route to reproduce the interacting DM-DE models without any phenomenological basis\footnote{Let us note that the interacting DM-DE models are very promising to alleviate the Hubble tension, see Tables B1 and B2 of \cite{DiValentino:2021izs}. }.

In this work we consider the cosmological dynamics in the Weyl Integrable geometry,  a generalization of the Einstein-Hilbert action. We note that the Weyl Integrable spacetime did not get much attention in the literature without any specific reason (only a few works discussing the astrophysical and cosmological consequences are available \cite{Salim:1999iz,Gannouji:2011va,Scholz:2014tba,Aguilar:2015tea,Romero:2015zmy,Paliathanasis:2020dbe,Paliathanasis:2020plf,Paliathanasis:2021bhj}). While it is interesting to notice that within this cosmological domain, one can mimic the late-time cosmic acceleration  \cite{Miritzis:2013ai,Aguila:2014moa}  and also the early inflationary phase  \cite{Fabris:1998fe}. Here we have explicitly found here, the Weyl Integrable geometry has a very novel feature which enlightens the cosmology of interacting DM-DE models by offering a field theoretical prescription. It is widely known that the theory of interacting dark matter - dark energy (known as interacting dark energy or coupled dark energy) has gained significant attention for solving several cosmological problems, such as the cosmic coincidence problem (see Refs. \cite{Huey:2004qv, Berger:2006db,delCampo:2006vv,delCampo:2008sr}), cosmological tensions (see Ref. \cite{DiValentino:2021izs}). Here we explicitly show that the Weyl Integrable geometry can offer a solid foundation to the theory of interacting dark energy. The interacting dark energy models are mostly governed by some phenomenological choices of the interaction function, which modifies the expansion history at the background and perturbation levels. Although attempts to construct the interaction function from the action formalisms have been considered in the literature \cite{Gleyzes:2015pma,vandeBruck:2015ida,Boehmer:2015kta,Boehmer:2015sha,DAmico:2016jbm,Pan:2020zza}, nevertheless, we are not done yet.
In the present work, we show that the dark sectors' interaction naturally arises in the context of
Weyl Integrable spacetime quite very naturally
without any mathematical complexities. This result is one of the exciting outcomes of this work. Having this novel feature, we performed the dynamical system analysis of the underlying cosmological scenario in the background of a nonflat Friedmann-Lema\^{i}tre-Robertson-Walker (FLRW) line element.
The assumption of the nonflat FLRW universe model is not an artist's imagination but rather the analyses of the cosmic microwave background anisotropies from Planck (considering the Pilk likelihood) favour a closed universe at several standard deviations \cite{Planck:2018vyg, DiValentino:2019qzk, Handley:2019tkm, DiValentino:2020hov}.

Interactive matter-scalar field schemes have been explored before where the conservation equations have the structure
\begin{equation}
\dot{\rho}_ {m} + 3H \left(\rho_ {m} + p_ {m} \right)= Q, \quad
\dot {\phi} \left [\ddot {\phi} +3 H \dot {\phi} + V'(\phi) \right] =  -Q, \label{interacting-scheme}
\end{equation}%
where a dot means derivative with respect to cosmic time $t$, comma derivative with respect to $\phi$, $ \rho_m $ is the energy density of matter, $\phi $ is the scalar field, $V(\phi ) $ its potential, $ Q $ is the interaction term, and $H=\dot{a}/a=\theta/3$ stands for the  \emph{ Hubble parameter} (which is a general measure of the isotropic rate of spatial expansion), where $a$ denotes the scale factor of the universe.

In Ref. \cite{Leon:2020pvt} such a scalar field cosmology with a generalized harmonic potential was investigated in flat and negatively curved Friedmann-Lema\^{i}tre-Robertson-Walker and Bianchi I metrics. An interaction between the scalar field and matter is considered in the form $Q= -\mu/2 \rho_m \dot{\phi}$. Asymptotic methods and averaging theory were used to obtain relevant information about the solution space. In this approach, the Hubble parameter played the role of a time-dependent perturbation parameter which controls the magnitude of the error between full-system and time-averaged solutions as it decreases. This approach shows that full and time-averaged systems have the same asymptotic behaviour. Numerical simulations were presented as evidence of such behaviour. Moreover, the asymptotic behaviour of the solutions is independent of the coupling function.

Interactive schemes like \eqref{interacting-scheme} naturally  appears in Scalar Tensor theory (STT) of gravity. Say, a general class of STTs, written in the so-called
Einstein frame (EF), which is given by \cite{Kaloper:1997sh}
\begin{align}&S_{EF}=\int_{M_4} d{ }^4 x \sqrt{|g|}\left\{\frac{1}{2} R-\frac{1}{2} g^{\mu
\nu}\nabla_\mu\phi\nabla_\nu\phi-V(\phi)+\chi(\phi)^{-2}
\mathcal{L}_{\text{matter}}(\mu,\nabla\mu,\chi(\phi)^{-1}g_{\alpha\beta})\right\},\label{eq1}
\end{align}
where $R$ is the curvature scalar, $\phi$ is the
scalar field,   $\nabla_\alpha$
is the covariant derivative, $V(\phi)$ is the quintessence self-interacting potential,
$\chi(\phi)^{-2}$ is the coupling function, $\mathcal{L}_{\text{matter}}$
is the matter Lagrangian, and $\mu$ is a collective name for the
matter degrees of freedom,  repeated indexes mean sum over
them. The energy-momentum tensor of matter is defined by
\begin{equation}T_{\alpha
\beta}=-\frac{2}{\sqrt{|g|}}\frac{\delta}{\delta g^{\alpha
\beta}}\left\{\sqrt{|g|}
 \chi^{-2}(\phi)\mathcal{L}(\mu,\nabla\mu,\chi^{-1}(\phi)g_{\alpha
 \beta})\right\}.\label{Tab}\end{equation}

By considering the conformal transformation $\overline{g}_{\alpha
\beta}=\chi(\phi)^{-1}g_{\alpha \beta}$, defining the Brans-Dicke (BD) coupling ``constant'' $\omega(\chi)$ in such way that
$d\phi=\pm \sqrt{\omega(\chi)+3/2}\chi^{-1} d\chi$ and recalling
$\overline{V}(\chi)=\chi^2 V(\phi(\chi))$, the action (\ref{eq1}) can be
written in the Jordan frame (JF) as  \cite{Coley:2003mj}
\begin{align}& S_{JF}=\int_{M_4} d{ }^4 x \sqrt{|\overline{g}|}\left\{\frac{1}{2}\chi \overline{R}-\frac{1}{2}\frac{\omega(\chi)}{\chi}(\overline{\nabla}\chi)^2-\overline{V}(\chi)
+\mathcal{L}_{\text{matter}}(\mu,\nabla\mu,\overline{g}_{\alpha
\beta})\right\}.\label{eq1JF}
\end{align}
Here the bar is used to denote geometrical objects defined with
respect to the metric $\overline{g}_{\alpha \beta}$.
In the STT given by (\ref{eq1JF}), the energy-momentum of the
matter fields,\begin{equation}\overline{T}_{\alpha
\beta}=-\frac{2}{\sqrt{|\overline{g}|}}\frac{\delta}{\delta \overline{g}^{\alpha
\beta}}\left\{\sqrt{|\overline{g}|}
\mathcal{L}(\mu,\nabla\mu,\overline{g}_{\alpha
 \beta})\right\}, \label{Tabprime}\end{equation} is separately conserved. That is
$\overline{\nabla}^\alpha \overline{T}_{\alpha \beta}=0$.  However, when is written in
the EF (\ref{eq1}), with a matter energy-momentum tensor given by \eqref{Tab}, this is no longer the case (although the
overall energy density is conserved). In fact in the EF we find
that
\begin{equation}
   Q_\beta\equiv\nabla^\alpha T_{\alpha \beta}=-\frac{1}{2}T {\chi(\phi)}^{-1}\frac{\mathrm{d}\chi(\phi)}{\mathrm{d}\phi}\nabla_{\beta}\phi,\quad
 T=T^\alpha_\alpha.
\end{equation}
 For action \eqref{eq1}, the strength of the coupling between the perfect fluid and the scalar field is
$Q=\frac{1}{2}(4-3\gamma)\rho_m\dot\phi
\frac{\mathrm{d}\ln \chi(\phi)}{\mathrm{d}\phi}$, where $ \chi(\phi)$ is an input function. In reference  \cite{Billyard:2000bh}  the interaction terms (in
the flat FLRW geometry) $Q=\alpha\dot\phi\rho_m$ and
$Q=\alpha\rho_m H$ were investigated, here $\alpha$ is a constant, $\phi$ is the
scalar field, $\rho$ is the energy density of background matter
and $H$ is the Hubble parameter. The first choice
corresponds to an exponential coupling function $\chi(\phi)=\chi_0
\exp\left(2 \alpha \phi/(4-3\gamma)\right).$ The second case
corresponds to the choice $\chi=\chi_0 a^{-2\alpha/(4-3\gamma)}$
(and then, $\rho\propto a^{\alpha-3\gamma}$). Other models within the general setup  \eqref{interacting-scheme}, incorporates effective interaction terms  $ Q = 3 \alpha H \rho_m, \; Q = 3 \beta H \rho_\phi $ and $ Q = 3 H (\alpha \rho_m + \beta \rho_\phi)$ \cite{Cardenas:2018nem,Lepe:2015qhq}.

The article has been organized as follows. In section \ref{sec21}, we provide the gravitational equations of the Weyl Integrable spacetime in the FLRW universe model. In section \ref{sec-dyn-analysis} we perform the dynamical system analysis of the cosmological scenario considering the spatial curvature of the universe. Finally, in section \ref{sec-conclusion} we describe the main findings of this article in brief.

\section{Weyl Integrable spacetime}
\label{sec21}

In Weyl Integrable Spacetime (WIST) or Weyl Integrable Geometry (WIG), the
Einstein-Hilbert Action is modified as
\begin{equation}
S_{W}=\int dx^{4}\sqrt{-g}\left(  \widetilde{R}+\xi\left(  \widetilde{\nabla}_{\nu
}\left(  \widetilde{\nabla}_{\mu}\phi\right)  \right)  g^{\mu\nu}-V\left(
\phi\right)  +\mathcal{L}_{m}\right)  ,
\end{equation}
in which $g_{\mu\nu}$ is the metric tensor for the physical space,
$\widetilde{\nabla}_{\mu}$ denotes covariant derivative defind by the symbols
$\widetilde{\Gamma}_{\mu\nu}^{\kappa}$, where $\widetilde{\Gamma}_{\mu\nu}^{\kappa}$
are the Christoffel symbols for the conformally related metric $\widetilde{g}%
_{\mu\nu}=\phi g_{\mu\nu}$. Parameter $\xi$ is an arbitrary
coupling constant, and $\mathcal{L}_{m}$ is the Lagrangian function for the
matter source.

When $\mathcal{L}_{m}$ describes a perfect fluid with energy density $\rho$
and pressure component $p$, the field equations in the Einstein-Weyl theory
are derived to be%
\begin{align}
& \widetilde{G}_{\mu\nu}+\widetilde{\nabla}_{\nu}\left(  \widetilde{\nabla}_{\mu}%
\phi\right)  -\left(  2\xi-1\right)  \left(  \widetilde{\nabla}_{\mu}\phi\right)
\left(  \widetilde{\nabla}_{\nu}\phi\right)  +\xi g_{\mu\nu}g^{\kappa\lambda
}\left(  \widetilde{\nabla}_{\kappa}\phi\right)  \left(  \widetilde{\nabla}_{\lambda
}\phi\right)  -V\left(  \phi\right)  g_{\mu\nu} \nonumber \\
& =-\left(  \widetilde{\rho}%
+\widetilde{p}\right)  u_{\mu}u_{\nu}-\widetilde{p}g_{\mu\nu}, \label{ww.08}%
\end{align}
where $\widetilde{G}_{\mu\nu}$ is the Einstein tensor with respect the metric
$\widetilde{g}_{\mu\nu}$ and $u^{\mu}$ is the comovig observer. The new parameters
$\widetilde{\rho},~\widetilde{p}\,\ $are the energy density and pressure~components
for the matter source multiplied by the factor $e^{-\frac{\phi}{2}}$, that means $(\widetilde{\rho},~\widetilde{p}) =  (e^{-\frac{\phi}{2}} \rho_m, e^{-\frac{\phi}{2}} p_m)$.

The field equations (\ref{ww.08}) can be written in the equivalent form
\begin{equation}
G_{\mu\nu}-\lambda\left(  \phi_{,\mu}\phi_{,\nu}-\frac{1}{2}g_{\mu\nu}%
\phi^{,\kappa}\phi_{,\kappa}\right)  -V\left(  \phi\right)  g_{\mu\nu
}=-\left(  \widetilde{\rho}+\widetilde{p}\right)  u_{\mu}u_{\nu}-\widetilde{p}g_{\mu\nu}
\label{ww.11}%
\end{equation}
where $G_{\mu\nu}$ is the Einstein tensor for the background space $g_{\mu\nu}$ and $\lambda=2\xi-\frac{3}{2}$.

Now, to proceed with the cosmological dynamics, we consider a homogeneous and isotropic universe characterized by the Friedmann-Lema\^{i}tre-Robertson-Walker (FLRW) line element
\begin{equation}
ds^{2}=-dt^{2}+a^{2}\left(  t\right)  \left[ \frac{dr^{2}}{1-Kr^{2}}%
+r^{2}\left(  d\theta^{2}+\sin^{2}\theta d\varphi^{2}\right] \right),
\label{ww.16}%
\end{equation}
where $a(t)$ is the expansion scale factor of the universe and $K$ is the curvature scalar of the universe which can take any of the values of $\{0, + 1, -1\}$.  For $K =0, +1, -1$, spatially flat, closed and open geometries of the universe are represented.
For the above FLRW metric, with lapse function $N=1$, one can write down the Friedmann equations for dust matter ($p_m=0$)  as follows
\begin{equation}
\left(  \frac{\theta^{2}}{3}+\frac{3K}{a^{2}}\right)  -\frac{\lambda}{2}\dot{\phi}^{2}-V\left(  \phi\right)  -e^{-\frac{\phi}{2}}\rho_m=0,
\label{ww.17}%
\end{equation}%
\begin{equation}
\dot{\theta} +\frac{\theta^2}{2}+ \frac{3K}{2a^{2}}-\frac{3}{2}V(\phi)+\frac{3}{4}\lambda\dot{\phi}^{2}=0, \label{ww.18}%
\end{equation}%
\begin{equation}
\ddot{\phi}+\theta\dot{\phi}+\frac{1}{\lambda}V_{,\phi}+\frac{1}{2\lambda}e^{-\frac{\phi}{2}}\rho_m=0,
\label{ww.19a}%
\end{equation}%
\begin{equation}
\dot{\rho_m}+\theta  \rho_m   -\rho_m\dot{\phi}=0. \label{ww.20a}%
\end{equation}
where we have considered the comoving observer $u^{\mu}=\delta
_{t}^{\mu}$, an overhead dot represents the cosmic time differentiation and $\theta=3\dot{a}/a$ is the expansion rate of the FLRW universe.  The conservation equations for the scalar field and the matter sector are of special importance in this context. The eqns. (\ref{ww.19a}) and (\ref{ww.20a}) can be alternatively expressed as
\begin{eqnarray}
&&\dot{\rho}_{\phi} + \theta (\rho_{\phi} + p_{\phi}) = - \frac{1}{2} \; \widetilde{\rho}\; \dot{\phi}, \label{eq9}\\
&&\dot{\widetilde{\rho}} + \theta (\widetilde{\rho} + \widetilde{p}) =  \frac{1}{2}\; \widetilde{\rho}\; \dot{\phi}, \label{eq10}
\end{eqnarray}
where $\rho_{\phi} = \frac{\lambda}{2}\;(\dot{\phi}^2 + V(\phi))$, $p_{\phi} = \frac{\lambda}{2}\; (\dot{\phi}^2 - V(\phi))$ and $\widetilde{p}=0$. One can clearly see that the above two equations
represent an interacting two-fluid system: $\nabla_{\nu} T^{\mu \nu}_{\phi} = - Q(t) = - \nabla_{\nu} T^{\mu \nu}_{\rm matter}$
where the interaction function $Q (t)$ is given by
\begin{equation}
\label{int_Q}
 Q (t)=   \frac{1}{2} \widetilde{\rho}(t)\; \dot{\phi}(t).
\end{equation}
Eqs. \eqref{eq9}-\eqref{eq10} realizes such an interactive scheme like \eqref{interacting-scheme}, with $p_m=0$ (dust matter). It is a very intriguing result in this context because the above formalism naturally induces an interaction between the scalar field and the matter field without mathematical complexities. We note that in General Relativity, usually one needs to introduce the interaction between the dark matter and dark energy by hand, but here in the WIST, one can see that an interaction term $Q$ is already present. It is one of the essences of WIST where we naturally have an interaction term between the scalar field and the matter sector. We further note that the above interaction may allow a changeable sign property depending on the sign of $\dot{\phi}$. We refer to Refs. \cite{Sun:2010vz,Wei:2010cs,Forte:2013fua,Guo:2017deu,Arevalo:2019axj,Pan:2019jqh} where the sign changeable interaction functions have been investigated from the phenomenological ground.

In the line of cosmological parameters,
the equation of state parameter for the effective fluid is defined to be
$w_{tot}\left(  t\right):= \frac{p_{tot}(t)}{\rho_{tot}(t)}$, and it is related to the
deceleration parameter $q(t)$, in the presence of curvature, by
\begin{equation}
q\left(  t\right):=-1-\frac{3 \dot{\theta}(t)}{\theta(t)^2}  =\frac{\left(  1+3w_{tot}\left(  t\right)  \right)  }%
{2}\left(  1+\frac{9 K}{a(t)\theta(t)^{2}}\right). \label{ww.21}%
\end{equation}

Since in this study we consider that the matter source is that of the dust fluid,
i.e $p_m=0$, from equation (\ref{ww.20a}), it follows $\rho_m=\rho_{m0}a^{-3}%
e^{\phi}$, thus we end with the set of differential equations (\ref{ww.17}),
(\ref{ww.18}) and (\ref{ww.19a}).

In the following section we present a detailed analysis of the dynamics of the
field equations (\ref{ww.17}), (\ref{ww.18}), (\ref{ww.19a}) and
(\ref{ww.20a}) by investigating the stationary points and their stability in new
dimensionless variables. Such an analysis is essential to understand further the effects of the curvature term in the cosmological dynamics of the WIST.

\section{Dynamical analysis}
\label{sec-dyn-analysis}

We consider the dimensionless dependent variables $\left\{  x,y,\omega
_{m},\eta,\Omega_K\right\}  $ defined as%
\begin{align}
& \dot{\phi}=\sqrt{2\left(  1+\frac{\theta^{2}}{3}\right)  }x, \\
& V\left(\phi\right)  =\left(  1+\frac{\theta^{2}}{3}\right)  y^{2}~, \\
& \rho_{m}=\left(
1+\frac{\theta^{2}}{3}\right)  e^{\frac{\phi}{2}}\Omega_m~, \label{ww.22}
\\
& K=\left(  1+\frac{\theta^{2}}{3}\right)  a^{2}\Omega_K, \\
& \theta=\frac
{\sqrt{3}\eta}{\sqrt{1-\eta^{2}} }, \label{ww.23}%
\end{align}
with inverse
\begin{align}
   & x= \frac{\dot{\phi}}{\sqrt{2} D},\\
   & y= \frac{\sqrt{V(\phi)}}{D}, \\
   & \Omega_m= \frac{e^{-\frac{\phi}{2}} \rho_m }{D^2},\\
   & \Omega_K=\frac{K}{a^2 D^2}, \quad D= \sqrt{1+\theta^2/3}\\
   & \eta= \frac{\theta/\sqrt{3}}{\sqrt{1+ \theta^2/3}},
\end{align}
and the new independent variable $dt=\sqrt{ 1+\frac{\theta^{2}}{3}}
d\tau$.

As $\eta \rightarrow \pm 1$ we have $\theta \rightarrow \pm \infty$.

In the new variables the field equations are described by the following
algebraic-differential system
\begin{equation}
\Omega_K-\frac{1}{3}\left(  \lambda x^{2}+y^{2}-\eta^{2}+\Omega_m\right)
=0 \label{ww.24}%
\end{equation}
with differential equations%
\begin{align}
& \frac{dx}{d\tau}=\frac{1}{12} \left(8 \sqrt{3} \eta  \lambda  x^3+2 \sqrt{3} \eta  x \left(2 \eta ^2-2
   y^2+ \Omega_{m}-6\right)-\frac{3 \sqrt{2} \left(2 \mu  y^2+ \Omega_{m}\right)}{\lambda }\right),  \label{ww.25}%
\\
& \frac{dy}{d\tau}=\frac{y}{6}  \left(\sqrt{3} \eta  \left(2 \eta ^2+4 \lambda  x^2-2 y^2+ \Omega_{m}\right)+3
   \sqrt{2} \mu  x\right),
\label{ww.26}%
\\
& \frac{d\Omega_m}{d\tau}=\frac{1}{6} \Omega_{m} \left(8 \sqrt{3} \eta  \lambda  x^2+3 \sqrt{2} x+2 \sqrt{3} \eta
   \left(2 \eta ^2-2 y^2+\Omega_{m}-3\right)\right), ~ \label{ww.27}%
\\
& \frac{d\eta}{d\tau}=\frac{\sqrt{3}}{6}\left(\eta ^2-1\right) \left(2 \eta ^2+4 \lambda  x^2-2 y^2+\Omega_{m}\right).
\label{ww.28}%
\end{align}
The new variable $\mu=V_{,\phi}/V$. In the following we shall consider
that $V\left(  \phi\right)  $ is the exponential potential such that
$\mu={\rm const}$.

We remark that under the discrete transformation $y\rightarrow-y$ the
dynamical system (\ref{ww.24})-(\ref{ww.28}) is invariant. Thus, without any loss
of generality we focus our analysis in the branch $y\geq0$, while by
definition $\Omega_m\geq0$ and $\eta^{2}\leq1$. Moreover, in the new
variables the effective equation of state parameter $w_{tot}$ and the
deceleration parameters read%
\begin{equation}
w_{tot}\left(  x,y,\Omega_m,\eta\right)  =\frac{\lambda  x^2-y^2}{\lambda  x^2+y^2+ \Omega_{m}}, \label{ww.29}%
\end{equation}
and%
\begin{equation}
q\left(  x,y,\Omega_m,\eta\right)  =\frac{4 \lambda  x^2-2 y^2+ \Omega_{m}}{2 \eta ^2}. \label{ww.30}%
\end{equation}

The stationary points of the dynamical system (\ref{ww.24})$-$(\ref{ww.28})
describe exact asymptotic solutions for the field equations. Indeed from
(\ref{ww.29}) we can determine explicitly the scale factor at every point
$P=\left(  x\left(  P\right)  ,y\left(  P\right)  ,\Omega_m\left(  P\right)
,\eta\left(  P\right)  \right)  $. Indeed for $w_{tot}\neq-1$, the exact
solution for the scale factor at a given point is $a\left(  t\right)
=a_{0}t^{\frac{2}{3\left(  1+w_{tot}\right)  }}$, while for $w_{tot}=-1$ we
calculate $a\left(  t\right)  =a_{0}e^{\theta_{0}t}$.

We proceed with our analysis to investigate the stationary points for
the field equations and study their stability properties. Such
analysis is essential to understand the evolution of the physical
variables and construct the cosmological history.

\subsection{Stationary points}

The stationary points for the dynamical system (\ref{ww.24})$-$(\ref{ww.28}) can
be categorized in the following families of points $P=\left(  x\left(
P\right)  ,y\left(  P\right)  ,\Omega_m\left(  P\right)  ,\eta\left(  P\right)
\right)  $.

\subsubsection{Points $P_{1}^{\pm}$}

The stationary points $P_{1}^{\pm}$ with coordinates
\begin{equation}
P_{1}^{\pm}=\left(  \frac{1}{\sqrt{\lambda}},0,0,\pm1\right)  .
\end{equation}
describe asymptotic solutions where only the kinetic component of the scalar
field potential contributes to the total cosmological solution. The stationary
points are real and physically accepted when $\lambda>0$. Moreover we derive
$\Omega_K\left(  P_{1}^{\pm}\right)  =0$, $q\left(  P_{1}^{\pm}\right)  =2$
and $w_{tot}\left(  P_{1}^{\pm}\right)  =1$. Hence the background space is
that of the spatially flat FLRW space with scale factor $a\left(  t\right)
=a_{0}t^{\frac{1}{3}}$.

In order to investigate the stability of the stationary points we derive the
eigenvalues of the following matrix around the stationary points%
\begin{equation}
\mathbf{A=}%
\begin{pmatrix}
\frac{\partial}{\partial x}\left(  \frac{dx}{d\tau}\right)  & \frac{\partial
}{\partial y}\left(  \frac{dx}{d\tau}\right)  & \frac{\partial}{\partial
\Omega_m}\left(  \frac{dx}{d\tau}\right)  & \frac{\partial}{\partial\eta
}\left(  \frac{dx}{d\tau}\right) \\
\frac{\partial}{\partial x}\left(  \frac{dy}{d\tau}\right)  & \frac{\partial
}{\partial y}\left(  \frac{dy}{d\tau}\right)  & \frac{\partial}{\partial
\Omega_m}\left(  \frac{dy}{d\tau}\right)  & \frac{\partial}{\partial\eta
}\left(  \frac{dy}{d\tau}\right) \\
\frac{\partial}{\partial x}\left(  \frac{d\Omega_m}{d\tau}\right)  &
\frac{\partial}{\partial y}\left(  \frac{d\Omega_m}{d\tau}\right)  &
\frac{\partial}{\partial\Omega_m}\left(  \frac{d\Omega_m}{d\tau}\right)  &
\frac{\partial}{\partial\eta}\left(  \frac{d\Omega_m}{d\tau}\right) \\
\frac{\partial}{\partial x}\left(  \frac{d\eta}{d\tau}\right)  &
\frac{\partial}{\partial y}\left(  \frac{d\eta}{d\tau}\right)  &
\frac{\partial}{\partial\Omega_m}\left(  \frac{d\eta}{d\tau}\right)  &
\frac{\partial}{\partial\eta}\left(  \frac{d\eta}{d\tau}\right)
\end{pmatrix}
.
\end{equation}

Indeed for the point $P_{1}^{+}$ the eigenvalues of the matrix $\mathbf{A}$
read%
\begin{equation}
e_{1}\left(  P_{1}^{+}\right)  =\frac{4}{\sqrt{3}}, e_{1}\left(  P_{1}%
^{+}\right)  =6\sqrt{3}, e_{3}\left(  P_{1}^{+}\right)  =\frac{\sqrt
{2}+2\sqrt{3\lambda}}{2\sqrt{\lambda}}, e_{4}\left(  P_{1}^{+}\right)
=\sqrt{3}+\frac{\mu}{\sqrt{2\lambda}}.
\end{equation}
Because $e_{1}\left(  P_{1}^{+}\right)  $, $e_{2}\left(  P_{1}^{+}\right)
\,\ $and $e_{3}\left(  P_{1}^{+}\right)  $ have always poisitive real parts,
the stationary point can not be an attractor which means that the asymptotic
solution at the point is unstable. Specifically for $\mu>-\sqrt{6\lambda}$,
$P_{1}^{+}$ is a source while for $\mu<\sqrt{6\lambda}$, $P_{1}^{+}$ is a
saddle point.

In a similar way the eigenvalues for the stationary point $P_{1}^{-}$ are%
\begin{equation}
e_{1}\left(  P_{1}^{-}\right)  =-\frac{4}{\sqrt{3}}, e_{1}\left(  P_{1}%
^{-}\right)  =-6\sqrt{3}, e_{3}\left(  P_{1}^{-}\right)  =\frac{\sqrt
{2}-2\sqrt{3\lambda}}{2\sqrt{\lambda}}, e_{4}\left(  P_{1}^{-}\right)
=-\sqrt{3}+\frac{\mu}{\sqrt{2\lambda}}.
\end{equation}
Thus for $\left\{  \lambda>\frac{1}{6},\mu\leq1\right\}  $ and $\left\{
\mu>0\text{, }\lambda>\frac{\mu^{2}}{6}\right\}  $ all the eigenvalues have
negative real parts from where we can infer that $P_{1}^{-}$ is an attractor
while the asymptotic solution at the specific point is a stable solution.

\subsubsection{Points $P_{2}^{\pm}$}

The family of points $P_{2}^{\pm}$ is consisted by the stationary points%
\begin{equation}
P_{2}^{\pm}=\left(  -\frac{1}{\sqrt{\lambda}},0,0,\pm1\right)  .
\end{equation}
The physical properties of the asymptotic solution at the points $P_{2}^{\pm}$ are similar to that of the points $P_{1}^{\pm}$; thus, we omit the discussion.
We proceed with the derivation of the eigenvalues for the linearized matrix
and the investigation of the stability properties.

Around the stationary point $P_{2}^{+}$ the eigenvalues of the linearized
system are
\begin{equation}
e_{1}\left(  P_{2}^{+}\right)  =\frac{4}{\sqrt{3}}, e_{2}\left(  P_{2}%
^{+}\right)  =6\sqrt{3}, e_{3}\left(  P_{2}^{+}\right)  =\frac{-\sqrt
{2}+2\sqrt{3\lambda}}{2\sqrt{\lambda}}, e_{4}\left(  P_{2}^{+}\right)
=\sqrt{3}-\frac{\mu}{\sqrt{2\lambda}},
\end{equation}
from where we infer that when $\,0<\lambda<\frac{1}{6}$ or $\mu>0,~0<\lambda
<\frac{\mu^{2}}{6}$ $\ $point $P_{1}^{+}$ is a saddle point, otherwise is a source.

For point $P_{2}^{-}$ we derive the eigenvalues%
\begin{equation}
e_{1}\left(  P_{2}^{-}\right)  =-\frac{4}{\sqrt{3}}, e_{2}\left(  P_{2}%
^{-}\right)  =-6\sqrt{3}, e_{3}\left(  P_{2}^{-}\right)  =-\frac{\sqrt
{2}+2\sqrt{3\lambda}}{2\sqrt{\lambda}}, e_{4}\left(  P_{2}^{-}\right)
=-\sqrt{3}-\frac{\mu}{\sqrt{2\lambda}},
\end{equation}
from where we conclude that when $\left\{  \lambda>0, \mu<0\right\}  $, or
$\left\{  \mu>0,\lambda>\frac{\mu^{2}}{6}\right\}  $ all the eigenvalues have
negative real parts from where we infer that point $P_{2}^{-}$ is an attractor.

\subsubsection{Points $P_{3}^{\pm}$}

Points $P_{3}^{\pm}$ with coordinates%
\begin{equation}
P_{3}^{\pm}=\left(  -\frac{\sqrt{6}}{\mu},\frac{\sqrt{2\left(  6\lambda
-\mu^{2}\right)  }}{\mu},0,\pm1\right)
\end{equation}
describe asymptotic exact solutions where both the kinetic part and the
potential terms contributes to the cosmological fluid. The points are
physically accepted when $6\lambda\geq\mu^{2}$, which means that $\lambda>0$.
The effective parameter for the equation of state is $w_{tot}\left(
P_{3}^{\pm}\right)  =1$, while the deceleration parameter is derived $q\left(
P_{3}^{\pm}\right)  =4\left(  -1+\frac{6\lambda}{\mu^{2}}\right)  $.
Furthermore, we derive $\Omega_K\left(  P_{3}^{\pm}\right)  =-1+\frac
{6\lambda}{\mu^{2}}$. Consequently, the stationary points when exist describe
exact solutions of FLRW universe with positive curvature, $\Omega_K\left(
P_{3}^{\pm}\right)  >0$, while always $q\left(  P_{3}^{\pm}\right)  >0$.

The corresponding eigenvalues are
\begin{align}
e_{1}\left(  P_{3}^{+}\right)   &  =6\sqrt{3}, e_{2}\left(  P_{3}^{+}\right)
=\sqrt{3}\frac{\left(  \mu-1\right)  }{\mu}, \\
e_{3}\left(  P_{3}^{+}\right)   &  =\frac{2}{\sqrt{3}}\left(  1-\frac
{\sqrt{2\left(  9\lambda-\mu^{2}\right)  }}{\mu}\right), ~e_{4}\left(
P_{3}^{+}\right)  =\frac{2}{\sqrt{3}}\left(  1+\frac{\sqrt{2\left(
9\lambda-\mu^{2}\right)  }}{\mu}\right)  ,
\end{align}
and%
\begin{align}
e_{1}\left(  P_{3}^{-}\right)   &  =-6\sqrt{3}, e_{2}\left(  P_{3}%
^{-}\right)  =-\sqrt{3}\frac{\left(  \mu-1\right)  }{\mu}, \\
e_{3}\left(  P_{3}^{-}\right)   &  =-\frac{2}{\sqrt{3}}\left(  1-\frac
{\sqrt{2\left(  9\lambda-\mu^{2}\right)  }}{\mu}\right), ~e_{4}\left(
P_{3}^{-}\right)  =-\frac{2}{\sqrt{3}}\left(  1+\frac{\sqrt{2\left(
9\lambda-\mu^{2}\right)  }}{\mu}\right)  ,
\end{align}

We conclude that points $P_{3}^{\pm}$ are always saddle points when they exist.

\subsubsection{Points $P_{4}^{\pm}$}

The stationary points which form the family of points $P_{4}^{\pm}$ have
coordinates%
\begin{equation}
P_{4}^{\pm}=\left(  \pm\sqrt{\frac{2}{3}},0,-\frac{8}{3}\lambda,\pm1\right)  .
\end{equation}
The points are physically accepted for $\lambda\,<0$. The asymptotic solutions
at the points describe FLRW spacetimes with curvature $\Omega_K\left(
P_{4}^{\pm}\right)  =-\frac{\left(  1+2\lambda\right)  }{3}$, that is
$-\frac{1}{2}<\lambda<0$, $\Omega_K\left(  P_{4}^{\pm}\right)  <0$ otherwise
$\Omega_K\left(  P_{4}^{\pm}\right)  >0$. The deceleration parameter and the
effective equation of state parameter are calculated $q\left(  P_{4}^{\pm
}\right)  =0$ and $w_{tot}\left(  P_{4}^{\pm}\right)  =-\frac{1}{3}$.

As far as the stability properties of the stationary points are concerned, we
derive the eigenvalues for the linearized system near to the points, they are%
\begin{equation}
e_{1}\left(  P_{4}^{+}\right)  =2\sqrt{3}, e_{2}\left(  P_{4}^{+}\right)
=\frac{1+\mu}{\sqrt{3}}, e_{3}\left(  P_{4}^{+}\right)  =-\frac
{1+\sqrt{3+4\lambda}}{\sqrt{3}}, e_{4}\left(  P_{4}^{+}\right)
=-\frac{1-\sqrt{3+4\lambda}}{\sqrt{3}},
\end{equation}
and%
\begin{equation}
e_{1}\left(  P_{4}^{-}\right)  =-2\sqrt{3}, e_{2}\left(  P_{4}^{-}\right)
=-\frac{1+\mu}{\sqrt{3}}, e_{3}\left(  P_{4}^{-}\right)  =\frac
{1+\sqrt{3+4\lambda}}{\sqrt{3}}, e_{4}\left(  P_{4}^{-}\right)
=\frac{1-\sqrt{3+4\lambda}}{\sqrt{3}}.
\end{equation}

It is straightforward to conclude that $P_{4}^{\pm}$ are saddle points.

\subsubsection{Points $P_{5}^{\pm}$}

Points $P_{5}^{\pm}$ have coordinates
\[
P_{5}^{\pm}=\left(  \mp\frac{1}{\sqrt{6}\lambda},0,1-\frac{1}{6\lambda}%
,\pm1\right)  .
\]
From the constraint $\Omega_m\geq0$, it follows that parameter $\lambda$ is
constraint as, $\lambda<0$ or $\lambda>\frac{1}{6}$ .\ The asymptotic solution
at the points describe spatially flat FLRW solutions, i.e. $\Omega_K\left(
P_{5}^{\pm}\right)  =0$, with $q\left(  P_{5}^{\pm}\right)  =\frac{1}%
{4}\left(  2+\frac{1}{\lambda}\right)  $ and $w_{tot}\left(  P_{5}^{\pm
}\right)  =\frac{1}{6\lambda}$. Hence, acceleration exists when $\,-\frac
{1}{2}<\lambda<0$.

The eigenvalues of the linearized system around the stationary points are
determined
\begin{equation}
e_{1}\left(  P_{5}^{+}\right)  =\frac{\sqrt{3}}{2\lambda}\left(
1+6\lambda\right), ~e_{2}\left(  P_{5}^{+}\right)  =\frac{1+2\lambda}%
{2\sqrt{3}\lambda}, e_{3}\left(  P_{5}^{+}\right)  =\frac{1-6\lambda}%
{4\sqrt{3}\lambda}, e_{4}\left(  P_{5}^{+}\right)  =\frac{1+6\lambda-2\mu
}{4\sqrt{3}\lambda},
\end{equation}
and%
\begin{equation}
e_{1}\left(  P_{5}^{-}\right)  =-\frac{\sqrt{3}}{2\lambda}\left(
1+6\lambda\right), ~e_{2}\left(  P_{5}^{-}\right)  =-\frac{1+2\lambda
}{2\sqrt{3}\lambda}, e_{3}\left(  P_{5}^{-}\right)  =-\frac{1-6\lambda
}{4\sqrt{3}\lambda}, e_{4}\left(  P_{5}^{-}\right)  =-\frac{1+6\lambda-2\mu
}{4\sqrt{3}\lambda}.
\end{equation}

Hence, point $P_{5}^{+}$ is an attractor when $\left\{  \mu\leq0,-\frac{1}%
{6}<\lambda<0\right\}  \,$, $\left\{  0<\mu<\frac{1}{2},\frac{2\mu-1}%
{6}<\lambda<0\right\}  $, while point $P_{5}^{-}\,\ $is always a saddle point.
We observe that point $P_{5}^{+}$ can describe an accelerated universe as a
future attractor for the dynamical system

\subsubsection{Points $P_{6}^{\pm}$}

The family of points $P_{6}^{\pm}$ has coordinates%
\[
P_{6}^{\pm}=\left(  \pm\frac{\sqrt{6}}{1-2\mu},\frac{\sqrt{\left(
1+6\lambda-2\mu\right)  \left(  1+\mu\right)  }}{1-2\mu},\frac{12\lambda
\left(  \mu-1\right)  }{\left(  1-2\mu\right)  ^{2}},\pm1\right)  .
\]
\newline

\begin{figure}[ptb]
\centering\includegraphics[width=1\textwidth]{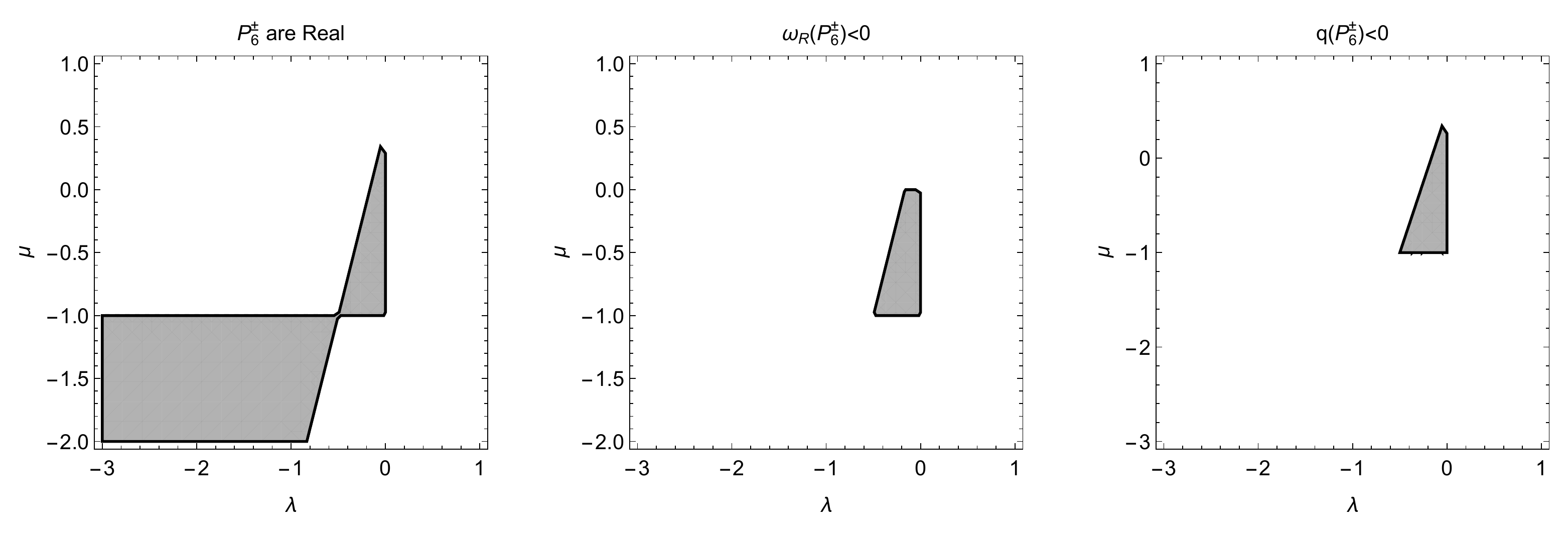}\caption{ Region plots
in the two-dimensional space for the variables $\left\{  \lambda,\mu\right\}
$, where the stationary points $P_{6}^{\pm}$ are real (left plot), the
spatially curvature is negative, $\Omega_K\left(  P_{6}^{\pm}\right)
<0~$(center plot), and the exact solution describes an accelerated universe
$~q\left(  P_{6}^{\pm}\right)  <0$ (right plot)}%
\label{fig01}%
\end{figure}

The points are real and physically accepted when the free parameters $\lambda$
and $\mu$ are constraint as $\left\{  \frac{1}{2}+3\lambda<\mu<-1\right\}  $
or $\left\{  \lambda<0,-1<\mu<\frac{1}{2}+3\lambda\right\}  \,$. The physical
parameters at the stationary points are derived $\Omega_K\left(  P_{6}^{\pm
}\right)  =\frac{\left(  1+6\lambda-2\mu\right)  \mu}{\left(  1-2\mu\right)
^{2}},~q\left(  P_{6}^{\pm}\right)  =\frac{\left(  1+\mu\right)  \left(
\mu-1-18\lambda\mu+2\mu^{2}\right)  }{\left(  1-2\mu\right)  ^{3}}$ and
$w_{tot}\left(  P_{6}^{\pm}\right)  =\frac{1}{2\mu-1}$. In Fig. \ref{fig01} we
present the region plots in the two-dimensional space for the variables
$\left\{  \lambda,\mu\right\}  $, where the stationary points $P_{6}^{\pm}$
are real, the spatially curvature is negative, i.e. $\Omega_K\left(
P_{6}^{\pm}\right)  <0$, and the exact solution describes an accelerated
universe $~q\left(  P_{6}^{\pm}\right)  <0$. We observe that when~ $\omega
_{R}\left(  P_{6}^{\pm}\right)  <0$, then the exact solution describes
acceleration.

\begin{figure}[ptb]
\centering\includegraphics[width=1\textwidth]{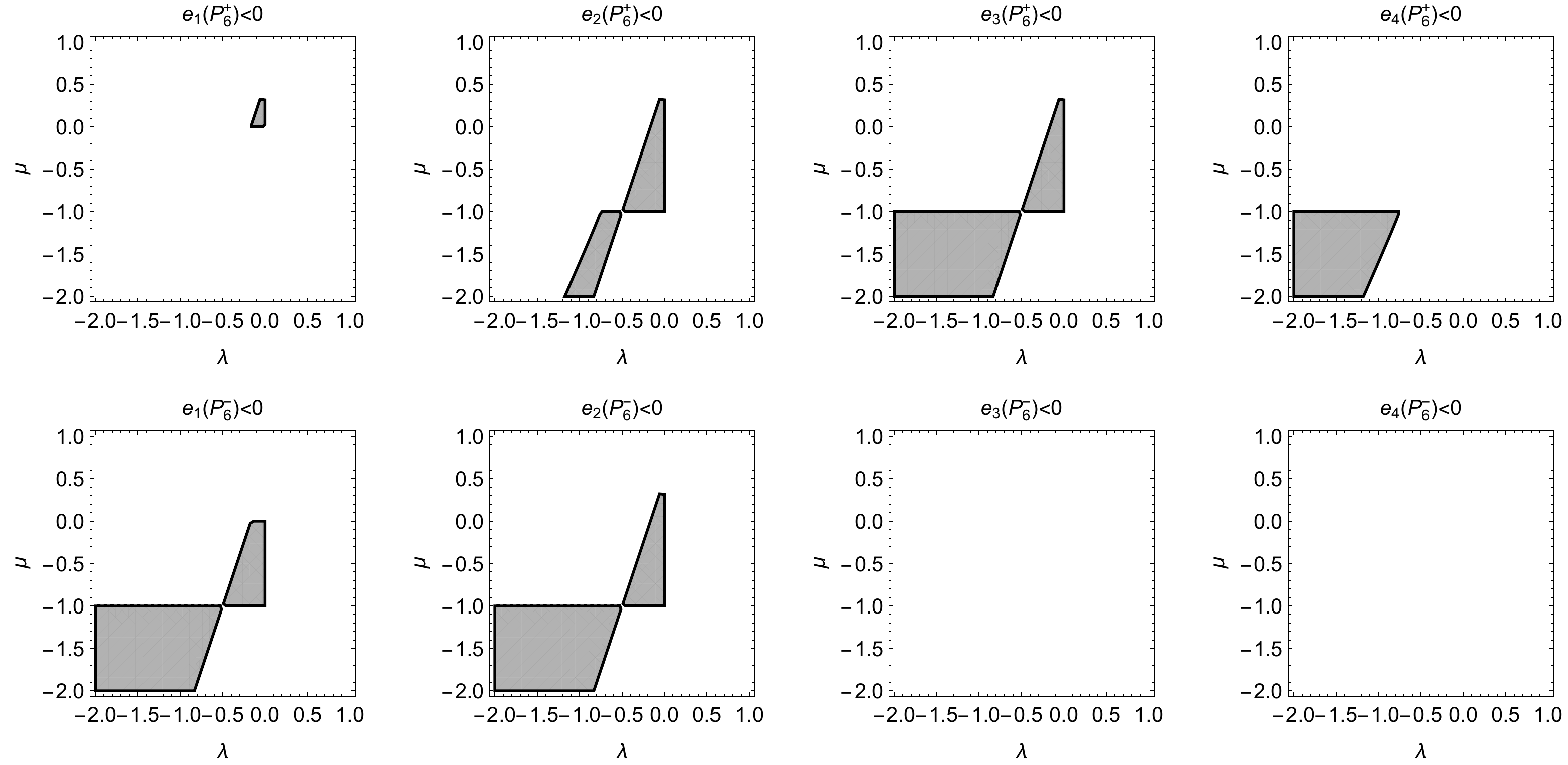}\caption{ Region plots
in the two-dimensional space for the variables $\left\{  \lambda,\mu\right\}
$, where the real parts of the eivenvalues for the linearized system at the
points $P_{6}^{\pm}$ are negatives. We observe that points $P_{6}^{\pm}$ are
always saddle points when they exist. }%
\label{fig02}%
\end{figure}

Furthermore, we determine the eigenvalues of the dynamical system numerically
around the stationary points $P_{6}^{\pm}$ and we found that the stationary
points $P_{6}^{\pm}$ are always saddle points when they exist. In Fig.
\ref{fig02} we present the regions in the space $\left\{  \lambda,\mu\right\}
$ in which the real part of the eigenvalues is negative.

\subsubsection{Points $P_{7}^{\pm}$}

Points $P_{7}^{\pm}$ are similar with $P_{6}^{\pm}$ where now
\[
P_{7}^{\pm}=\left(  \pm\frac{\sqrt{6}}{1-2\mu},-\frac{\sqrt{\left(
1+6\lambda-2\mu\right)  \left(  1+\mu\right)  }}{1-2\mu},\frac{12\lambda
\left(  \mu-1\right)  }{\left(  1-2\mu\right)  ^{2}},\pm1\right)  .
\]
points are physically accepted when $\left\{  1<\mu<\frac{1}{2}+3\lambda
,~\lambda>\frac{1}{6}\right\}  $. The physical parameters $\Omega_K$, $q$
and $w_{tot}$ have similar functional form with that of points $P_{6}^{\pm}$.
However in contrary with before, when the points exist, the asymptotic
solutions describe FLRW universes with positive spatial curvature and no acceleration.

\begin{figure}[ptb]
\centering\includegraphics[width=1\textwidth]{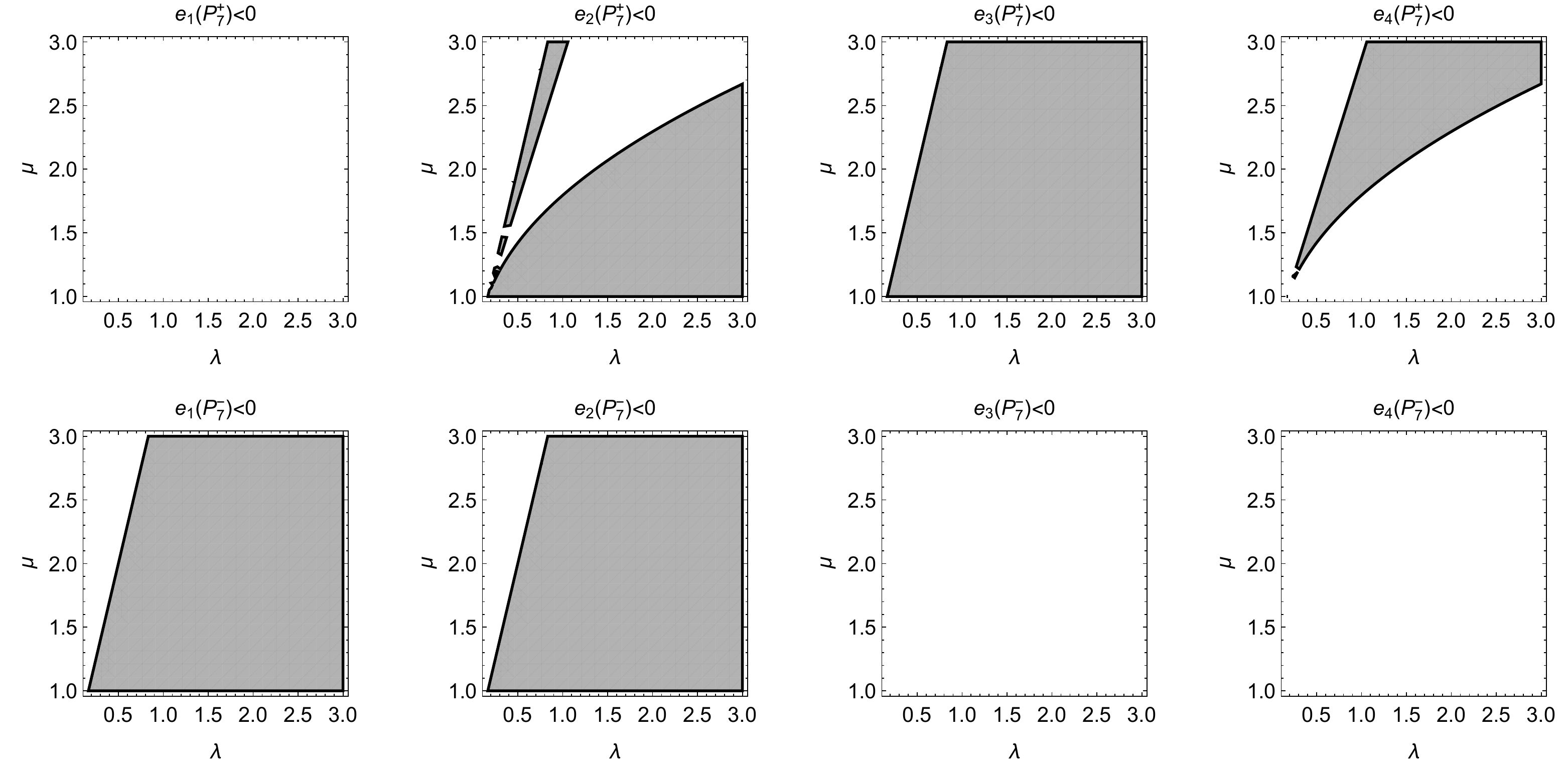}\caption{ Region plots
in the two-dimensional space for the variables $\left\{  \lambda,\mu\right\}
$, where the real parts of the eivenvalues for the linearized system at the
points $P_{7}^{\pm}$ are negatives. We observe that points $P_{7}^{\pm}$ are
always saddle points. }%
\label{fig03}%
\end{figure}

In Fig. \ref{fig03} we present the regions in the space $\left\{  \lambda, \mu\right\}  $ where the eigenvalues of the linearized system around the
points $P_{7}^{\pm}$ have negative real components. We conclude that the
stationary points $P_{7}^{\pm}$ when they exist, are always saddle points.

\subsubsection{Points $P_{8}^{\pm}$}

The set of stationary points $P_{8}^{\pm}$ with coordinates%
\begin{equation}
P_{8}^{\pm}=\left(  0,0,0,\pm1\right)  ,
\end{equation}
describes the Milne solution, $\Omega_K\left(  P_{8}^{\pm}\right)
=-\frac{1}{3}$, $q\left(  P_{8}^{\pm}\right)  =0$ and $w_{tot}\left(
P_{8}^{\pm}\right)  =-\frac{1}{3}$.

The eigenvalues of the linearized system around $P_{8}^{\pm}$ are
\begin{equation}
e_{1}\left(  P_{8}^{+}\right)  =2\sqrt{3}, e_{2}\left(  P_{8}^{+}\right)
=-\frac{2}{\sqrt{3}}, e_{3}\left(  P_{8}^{+}\right)  =-\frac{1}{\sqrt{3}%
}\text{ },~e_{4}\left(  P_{8}^{+}\right)  =\frac{1}{\sqrt{3}}%
\end{equation}
and%
\begin{equation}
e_{1}\left(  P_{8}^{-}\right)  =-2\sqrt{3}, e_{2}\left(  P_{8}^{-}\right)
=\frac{2}{\sqrt{3}}, e_{3}\left(  P_{8}^{-}\right)  =\frac{1}{\sqrt{3}}\text{
},~e_{4}\left(  P_{8}^{-}\right)  =-\frac{1}{\sqrt{3}}%
\end{equation}
that is, the Milne solutions are unstable solutions, since points $P_{8}^{\pm
}$ are always saddle points.

\subsubsection{Point $P_{9}$}

Point $P_{9}=\left(  0,0,0,0\right)  $, describes the Minkowski spacetime,
$\Omega_{K}\left(  P_{9}\right)  =0$. From numerical simulations it is easy to
observe that the Minkowski vacuum solution is always unstable.

\subsubsection{Line of points $P_{10}$}

Finally, the line of points $P_{10}$,%
\begin{equation}
P_{10}=\left(  0,\eta,0,\eta\right)
\end{equation}
describe de Sitter universes, $\Omega_{K}\left(  P_{10}\right)  =0$,
~$q\left(  P_{10}\right)  =-1$ and $w_{tot}\left(  P_{10}\right)  =-1.$ At the
asymptotic solution only the scalar field potential contributes to the
cosmological evolution.

The eigenvalues of the linearized system are calculated%
\[
e_{1}\left(  P_{10}\right)  =0,e_{2}\left(  P_{10}\right)  =-\sqrt{3}%
\eta,e_{3}\left(  P_{10}\right)  =-\sqrt{3}\eta,e_{4}\left(  P_{10}\right)
=\frac{2\eta\left(  2\eta^{2}-3\right)  }{\sqrt{3}}.
\]
Thus for $\eta<0$, the points always describe unstable de Sitter universes
(saddle or local source), while for $0<\eta<\sqrt{3/2}$, there exists a
submanifold in which the stationary point describes the de Sitter universe as
a future attractor. The derivation of the submanifold is a mathematical
calculation which does not contribute to the physical discussion. Hence, it is
omitted. The cases $1<\eta<\sqrt{3/2}$ and $\eta\geq\sqrt{3/2}$ are not allowed
due to the existence condition $-1\leq\eta\leq1$.

\section{Conclusions}
\label{sec-conclusion}

The discrepancies in the cosmological data and the universe's curvature are two major issues in cosmology at present. The first one indicates the need
of revision of the standard cosmological model $\Lambda$CDM, and the second issue directly questions the geometry of the universe by pointing toward a preference for a closed universe at several standard deviations
\cite{Planck:2018vyg,DiValentino:2019qzk,Handley:2019tkm,DiValentino:2020hov}.

The revision of the $\Lambda$CDM model has been performed in various ways, and modification of Einstein's gravitational theory is one of them. Even though a cluster of cosmological models has been proposed in the literature, none of them can explain all the observational discrepancies. Thus,
from this ground, all cosmological models are equivalent. In the present article, we have considered a modification of Einstein's gravity, namely, the Weyl Integrable geometry and explored the cosmological scenarios in the presence of the nonflat FLRW line element as this describes the most
generalized cosmic structure.

We have found some exciting results from this modified gravitational
theory. In particular, we found that the dark sector's interaction naturally
arises in this context without any mathematical complexities. That fills a
gap in the literature on interacting cosmology, which is mainly driven by the
phenomenological choices of the interaction function. In order to understand
the nature of the cosmological scenario derived from this gravitational
framework, we performed the dynamical system analysis and found a variety of possibilities.

The admitted stationary points of the dynamical system corresponding to ten
families of points, which can describe all the possible FLRW universes with
zero and non-zero spatial curvature. We conclude that the future attractors
of the dynamical system describe spatially flat FLRW geometries. Moreover,
exact solutions with acceleration are recovered. We conclude that Weyl
Integrable geometry can solve the flatness problem. Additionally, this is one
of the first studies in literature where interacting models in the dark sector
are discussed in the presence of curvature.

In future, we plan to investigate further this cosmological scenario by studying the evolution of the perturbations near the  stationary points
for the background geometry.

\acknowledgements

The research of AG was funded by Agencia Nacional de Investigaci\'{o}n y
Desarrollo - ANID through the program FONDECYT Regular grant no. 1200293. GL was funded by  Vicerrectoría de Investigación y Desarrollo Tecnológico at UCN.
 AP was partially supported by the National Research Foundation of South Africa
(Grant Numbers 131604). AP thanks Prof. G. Michalakopoulos and the Ionian University for the hospitality provided while part of this work carried out. SP acknowledges the financial support from the Department of Science and Technology (DST), Govt. of India, under the Scheme
``Fund for Improvement of S\&T Infrastructure (FIST)'' [File No. SR/FST/MS-I/2019/41].

\bibliographystyle{plain}
\bibliography{draft_curv}

\begin{thebibliography}{10}

\bibitem{DES:2017myr}
T.~M.~C. Abbott et~al.
\newblock {Dark Energy Survey year 1 results: Cosmological constraints from
  galaxy clustering and weak lensing}.
\newblock {\em Phys. Rev. D}, 98(4):043526, 2018.

\bibitem{Abdalla:2022yfr}
Elcio Abdalla et~al.
\newblock {Cosmology intertwined: A review of the particle physics,
  astrophysics, and cosmology associated with the cosmological tensions and
  anomalies}.
\newblock {\em JHEAp}, 34:49--211, 2022.

\bibitem{Planck:2018vyg}
N.~Aghanim et~al.
\newblock {Planck 2018 results. VI. Cosmological parameters}.
\newblock {\em Astron. Astrophys.}, 641:A6, 2020.
\newblock [Erratum: Astron.Astrophys. 652, C4 (2021)].

\bibitem{Aguila:2014moa}
Ricardo Aguila, Jos\'e~Edgar Madriz~Aguilar, Claudia Moreno, and Mauricio
  Bellini.
\newblock {Present accelerated expansion of the universe from new
  Weyl-Integrable gravity approach}.
\newblock {\em Eur. Phys. J. C}, 74(11):3158, 2014.

\bibitem{Aguilar:2015tea}
J.~E.~Madriz Aguilar, C.~Romero, J.~B. Fonseca~Neto, T.~S. Almeida, and J.~B.
  Formiga.
\newblock {(2+1)-Dimensional Gravity in Weyl Integrable Spacetime}.
\newblock {\em Class. Quant. Grav.}, 32(21):215003, 2015.

\bibitem{Arendse:2019hev}
Nikki Arendse et~al.
\newblock {Cosmic dissonance: are new physics or systematics behind a short
  sound horizon?}
\newblock {\em Astron. Astrophys.}, 639:A57, 2020.

\bibitem{Arevalo:2019axj}
Fabiola Arevalo, Antonella Cid, Luis~P. Chimento, and Patricio Mella.
\newblock {On sign-changeable interaction in FLRW cosmology}.
\newblock {\em Eur. Phys. J. C}, 79(4):355, 2019.

\bibitem{Berger:2006db}
Micheal~S. Berger and Hamed Shojaei.
\newblock {Interacting dark energy and the cosmic coincidence problem}.
\newblock {\em Phys. Rev. D}, 73:083528, 2006.

\bibitem{Billyard:2000bh}
Andrew~P. Billyard and Alan~A. Coley.
\newblock {Interactions in scalar field cosmology}.
\newblock {\em Phys. Rev. D}, 61:083503, 2000.

\bibitem{Boehmer:2015kta}
Christian~G. Boehmer, Nicola Tamanini, and Matthew Wright.
\newblock {Interacting quintessence from a variational approach Part I:
  algebraic couplings}.
\newblock {\em Phys. Rev. D}, 91(12):123002, 2015.

\bibitem{Boehmer:2015sha}
Christian~G. Boehmer, Nicola Tamanini, and Matthew Wright.
\newblock {Interacting quintessence from a variational approach Part II:
  derivative couplings}.
\newblock {\em Phys. Rev. D}, 91(12):123003, 2015.

\bibitem{Cai:2015emx}
Yi-Fu Cai, Salvatore Capozziello, Mariafelicia De~Laurentis, and Emmanuel~N.
  Saridakis.
\newblock {f(T) teleparallel gravity and cosmology}.
\newblock {\em Rept. Prog. Phys.}, 79(10):106901, 2016.

\bibitem{Cardenas:2018nem}
V\'\i{}ctor~H. C\'ardenas, Daniela Grand\'on, and Samuel Lepe.
\newblock {Dark energy and Dark matter interaction in light of the second law
  of thermodynamics}.
\newblock {\em Eur. Phys. J. C}, 79(4):357, 2019.

\bibitem{Clifton:2011jh}
Timothy Clifton, Pedro~G. Ferreira, Antonio Padilla, and Constantinos Skordis.
\newblock {Modified Gravity and Cosmology}.
\newblock {\em Phys. Rept.}, 513:1--189, 2012.

\bibitem{Coley:2003mj}
A.~A. Coley.
\newblock {\em {Dynamical systems and cosmology}}.
\newblock Kluwer, Dordrecht, Netherlands, 2003.

\bibitem{Copeland:2006wr}
Edmund~J. Copeland, M.~Sami, and Shinji Tsujikawa.
\newblock {Dynamics of dark energy}.
\newblock {\em Int. J. Mod. Phys. D}, 15:1753--1936, 2006.

\bibitem{DAmico:2016jbm}
Guido D'Amico, Teresa Hamill, and Nemanja Kaloper.
\newblock {Quantum field theory of interacting dark matter and dark energy:
  Dark monodromies}.
\newblock {\em Phys. Rev. D}, 94(10):103526, 2016.

\bibitem{DeFelice:2010aj}
Antonio De~Felice and Shinji Tsujikawa.
\newblock {f(R) theories}.
\newblock {\em Living Rev. Rel.}, 13:3, 2010.

\bibitem{delCampo:2006vv}
Sergio del Campo, Ramon Herrera, German Olivares, and Diego Pavon.
\newblock {Interacting models of soft coincidence}.
\newblock {\em Phys. Rev. D}, 74:023501, 2006.

\bibitem{delCampo:2008sr}
Sergio del Campo, Ramon Herrera, and Diego Pavon.
\newblock {Toward a solution of the coincidence problem}.
\newblock {\em Phys. Rev. D}, 78:021302, 2008.

\bibitem{DiValentino:2020vvd}
Eleonora Di~Valentino et~al.
\newblock {Cosmology intertwined III: $f\sigma_8$ and $S_8$}.
\newblock {\em Astropart. Phys.}, 131:102604, 2021.

\bibitem{DiValentino:2020zio}
Eleonora Di~Valentino et~al.
\newblock {Snowmass2021 - Letter of interest cosmology intertwined II: The
  hubble constant tension}.
\newblock {\em Astropart. Phys.}, 131:102605, 2021.

\bibitem{DiValentino:2019qzk}
Eleonora Di~Valentino, Alessandro Melchiorri, and Joseph Silk.
\newblock {Planck evidence for a closed Universe and a possible crisis for
  cosmology}.
\newblock {\em Nature Astron.}, 4(2):196--203, 2019.

\bibitem{DiValentino:2020hov}
Eleonora Di~Valentino, Alessandro Melchiorri, and Joseph Silk.
\newblock {Investigating Cosmic Discordance}.
\newblock {\em Astrophys. J. Lett.}, 908(1):L9, 2021.

\bibitem{DiValentino:2021izs}
Eleonora Di~Valentino, Olga Mena, Supriya Pan, Luca Visinelli, Weiqiang Yang,
  Alessandro Melchiorri, David~F. Mota, Adam~G. Riess, and Joseph Silk.
\newblock {In the realm of the Hubble tension\textemdash{}a review of
  solutions}.
\newblock {\em Class. Quant. Grav.}, 38(15):153001, 2021.

\bibitem{Fabris:1998fe}
J.~C. Fabris, J.~M. Salim, and S.~L. Sautu.
\newblock {Inflationary cosmological solutions in Weyl integrable geometry}.
\newblock {\em Mod. Phys. Lett. A}, 13:953--959, 1998.

\bibitem{Forte:2013fua}
M\'onica Forte.
\newblock {On extended sign-changeable interactions in the dark sector}.
\newblock {\em Gen. Rel. Grav.}, 46(10):1811, 2014.

\bibitem{Gannouji:2011va}
Radouane Gannouji, Hemwati Nandan, and Naresh Dadhich.
\newblock {FLRW cosmology in Weyl-Integrable Space-Time}.
\newblock {\em JCAP}, 11:051, 2011.

\bibitem{Gleyzes:2015pma}
J\'er\^ome Gleyzes, David Langlois, Michele Mancarella, and Filippo Vernizzi.
\newblock {Effective Theory of Interacting Dark Energy}.
\newblock {\em JCAP}, 08:054, 2015.

\bibitem{Guo:2017deu}
Juan-Juan Guo, Jing-Fei Zhang, Yun-He Li, Dong-Ze He, and Xin Zhang.
\newblock {Probing the sign-changeable interaction between dark energy and dark
  matter with current observations}.
\newblock {\em Sci. China Phys. Mech. Astron.}, 61(3):030011, 2018.

\bibitem{Handley:2019tkm}
Will Handley.
\newblock {Curvature tension: evidence for a closed universe}.
\newblock {\em Phys. Rev. D}, 103(4):L041301, 2021.

\bibitem{Hildebrandt:2016iqg}
H.~Hildebrandt et~al.
\newblock {KiDS-450: Cosmological parameter constraints from tomographic weak
  gravitational lensing}.
\newblock {\em Mon. Not. Roy. Astron. Soc.}, 465:1454, 2017.

\bibitem{Hildebrandt:2018yau}
H.~Hildebrandt et~al.
\newblock {KiDS+VIKING-450: Cosmic shear tomography with optical and infrared
  data}.
\newblock {\em Astron. Astrophys.}, 633:A69, 2020.

\bibitem{Huey:2004qv}
Greg Huey and Benjamin~D. Wandelt.
\newblock {Interacting quintessence. The Coincidence problem and cosmic
  acceleration}.
\newblock {\em Phys. Rev. D}, 74:023519, 2006.

\bibitem{Joudaki:2017zdt}
Shahab Joudaki et~al.
\newblock {KiDS-450 + 2dFLenS: Cosmological parameter constraints from weak
  gravitational lensing tomography and overlapping redshift-space galaxy
  clustering}.
\newblock {\em Mon. Not. Roy. Astron. Soc.}, 474(4):4894--4924, 2018.

\bibitem{Kaloper:1997sh}
Nemanja Kaloper and Keith~A. Olive.
\newblock {Singularities in scalar tensor cosmologies}.
\newblock {\em Phys. Rev. D}, 57:811--822, 1998.

\bibitem{Leon:2020pvt}
Genly Leon, Esteban Gonz\'alez, Alfredo~D. Millano, and Felipe Orlando~Franz
  Silva.
\newblock {A Perturbative Analysis of Interacting Scalar Field Cosmologies}, 3
  2020.

\bibitem{Lepe:2015qhq}
Samuel Lepe and Francisco Pe\~na.
\newblock {Interacting cosmic fluids and phase transitions under a holographic
  modeling for dark energy}.
\newblock {\em Eur. Phys. J. C}, 76(9):507, 2016.

\bibitem{Miritzis:2013ai}
John Miritzis.
\newblock {Acceleration in Weyl integrable spacetime}.
\newblock {\em Int. J. Mod. Phys. D}, 22:1350019, 2013.

\bibitem{Nojiri:2017ncd}
S.~Nojiri, S.D. Odintsov, and V.K. Oikonomou.
\newblock {Modified Gravity Theories on a Nutshell: Inflation, Bounce and
  Late-time Evolution}.
\newblock {\em Phys. Rept.}, 692:1--104, 2017.

\bibitem{Nojiri:2006ri}
Shin'ichi Nojiri and Sergei~D. Odintsov.
\newblock {Introduction to modified gravity and gravitational alternative for
  dark energy}.
\newblock {\em eConf}, C0602061:06, 2006.

\bibitem{Paliathanasis:2021bhj}
Andronikos Paliathanasis.
\newblock {Dynamical Analysis and Cosmological Evolution in Weyl Integrable
  Gravity}.
\newblock {\em Universe}, 7(12):468, 2021.

\bibitem{Paliathanasis:2020plf}
Andronikos Paliathanasis, Genly Leon, and John~D. Barrow.
\newblock {Einstein-aether theory in Weyl integrable geometry}.
\newblock {\em Eur. Phys. J. C}, 80(12):1099, 2020.

\bibitem{Paliathanasis:2020dbe}
Andronikos Paliathanasis, Genly Leon, and John~D. Barrow.
\newblock {Inhomogeneous spacetimes in Weyl integrable geometry with matter
  source}.
\newblock {\em Eur. Phys. J. C}, 80(8):731, 2020.

\bibitem{Pan:2020zza}
Supriya Pan, German~S. Sharov, and Weiqiang Yang.
\newblock {Field theoretic interpretations of interacting dark energy scenarios
  and recent observations}.
\newblock {\em Phys. Rev. D}, 101(10):103533, 2020.

\bibitem{Pan:2019jqh}
Supriya Pan, Weiqiang Yang, Chiranjeeb Singha, and Emmanuel~N. Saridakis.
\newblock {Observational constraints on sign-changeable interaction models and
  alleviation of the $H_0$ tension}.
\newblock {\em Phys. Rev.}, D100(8):083539, 2019.

\bibitem{Perivolaropoulos:2021jda}
Leandros Perivolaropoulos and Foteini Skara.
\newblock {Challenges for $\Lambda$CDM: An update}.
\newblock 5 2021.

\bibitem{SupernovaCosmologyProject:1996grv}
S.~Perlmutter et~al.
\newblock {Measurements of the cosmological parameters Omega and Lambda from
  the first 7 supernovae at z\ensuremath{>}=0.35}.
\newblock {\em Astrophys. J.}, 483:565, 1997.

\bibitem{SupernovaSearchTeam:1998fmf}
Adam~G. Riess et~al.
\newblock {Observational evidence from supernovae for an accelerating universe
  and a cosmological constant}.
\newblock {\em Astron. J.}, 116:1009--1038, 1998.

\bibitem{Riess:2021jrx}
Adam~G. Riess et~al.
\newblock {A Comprehensive Measurement of the Local Value of the Hubble
  Constant with 1 km/s/Mpc Uncertainty from the Hubble Space Telescope and the
  SH0ES Team}.
\newblock 12 2021.

\bibitem{Romero:2015zmy}
Jes\'us~Mart\'\i{}n Romero, Mauricio Bellini, and Jos\'e~Edgar Madriz~Aguilar.
\newblock {Gravitational waves from a Weyl-Integrable manifold: a new
  formalism}.
\newblock {\em Phys. Dark Univ.}, 13:1--6, 2016.

\bibitem{Salim:1999iz}
J.~M. Salim and S.~L. Sautu.
\newblock {Gravitational collapse in Weyl integrable space-times}.
\newblock {\em Class. Quant. Grav.}, 16:3281--3295, 1999.

\bibitem{Scholz:2014tba}
Erhard Scholz.
\newblock {MOND-like acceleration in integrable Weyl geometric gravity}.
\newblock {\em Found. Phys.}, 46(2):176--208, 2016.

\bibitem{Sotiriou:2008rp}
Thomas~P. Sotiriou and Valerio Faraoni.
\newblock {f(R) Theories Of Gravity}.
\newblock {\em Rev. Mod. Phys.}, 82:451--497, 2010.

\bibitem{Sun:2010vz}
Cheng-Yi Sun and Rui-Hong Yue.
\newblock {New Interaction between Dark Energy and Dark Matter Changes Sign
  during Cosmological Evolution}.
\newblock {\em Phys. Rev. D}, 85:043010, 2012.

\bibitem{vandeBruck:2015ida}
C.~van~de Bruck and J.~Morrice.
\newblock {Disformal couplings and the dark sector of the universe}.
\newblock {\em JCAP}, 04:036, 2015.

\bibitem{Wei:2010cs}
Hao Wei.
\newblock {Cosmological Constraints on the Sign-Changeable Interactions}.
\newblock {\em Commun. Theor. Phys.}, 56:972--980, 2011.

\bibitem{Weinberg:1988cp}
Steven Weinberg.
\newblock {The Cosmological Constant Problem}.
\newblock {\em Rev. Mod. Phys.}, 61:1--23, 1989.

\bibitem{Zlatev:1998tr}
Ivaylo Zlatev, Li-Min Wang, and Paul~J. Steinhardt.
\newblock {Quintessence, cosmic coincidence, and the cosmological constant}.
\newblock {\em Phys. Rev. Lett.}, 82:896--899, 1999.

\end{thebibliography}

\end{document}